# RADIO OBSERVATIONS OF THE HUBBLE DEEP FIELD SOUTH - A NEW CLASS OF RADIO-LUMINOUS GALAXIES?

R. P. NORRIS, A. HOPKINS, R. J. SAULT, R. D. EKERS, J. EKERS, F. BADIA, J. HIGDON, M. H. WIERINGA,
CSIRO Australia Telescope National Facility
P. O. Box 76, Epping, NSW 1710, Australia
E-mail: ray.norris@atnf.csiro.au

B. J. BOYLE
Anglo-Australian Observatory
P.O. Box 296, Epping, NSW 1710, Australia

R. E. WILLIAMS
Space Telescope Science Institute
3700, San Martin Drive, Johns Hopkins University Homewood Campus, Baltimore, MD 21218 USA

We present the first results from a series of radio observations of the Hubble Deep Field South and its flanking fields. Here we consider only those sources greater than 100 µJy at 20 cm, in an 8-arcmin square field that covers the WFPC field, the STIS and NICMOS field, and most of the HST flanking fields and complementary ground-based observations. We have detected 13 such radio sources, two of which are in the WFPC2 field itself. One of the sources in the WFPC field (source c) corresponds to a very faint galaxy, and several others outside the WFPC field can not be identified with sources in the other optical/IR. The radio and optical luminosities of these galaxies are inconsistent with either conventional starburst galaxies or with radio-loud galaxies. Instead, it appears that it belongs to a population of galaxies which are rare in the local Universe, possibly consisting of a radio-luminous active nucleus embedded in a very dusty starburst galaxy, and which are characterised by a very high radio/optical luminosity ratio.

## 1   Introduction

In December 1995 the WFPC2 camera of the Hubble Space Telescope (HST) observed a 2-arcmin patch of sky for 10 days, producing the deepest high-resolution image ever of the sky. The resulting images (the Hubble Deep Field North, or HDF-N) contain about 1500 galaxies, down to 30th magnitude, and give a unique slice through the early evolution of the Universe. These data provide powerful tests for models of early galaxy formation. Other projects being tackled include galaxy evolution, clustering studies, and cosmological modelling. Extremely deep radio images of this field have been made with the Very Large Array [1,2] and most of the sources found in these radio observations are starburst galaxies at a redshift of typically 0.5.

A corresponding deep field in the Southern Hemisphere (HDF-S) was observed in October 1998 [3]. Criteria for selecting this field included not only HST constraints but also constraints such as low far-infrared cirrus, low neutral hydrogen column density, etc. They also include the requirement that there should be no strong radio source in the field. (< 1 mJy at 3 cm), so that a deep radio image could also be obtained of the field, and the Australia Telescope Compact Array (ATCA) was used for this selection process. The HDF-S differs from the HDF-N in a number of respects:
- a field was chosen which has a nearby quasar, to enable studies of the intervening intergalactic medium
- greater attention was paid to the area (the "flanking fields") surrounding the primary WFPC field
- many ground-based observatories mounted simultaneous efforts to study the HDF-S, and most of these data were released into the public domain in advance of publication.

Unfortunately, the only suitable quasar was only 10 arcmin from the second strongest (140 mJy) radio source in any of the candidate fields searched. Nevertheless, this field has now been selected for the Southern Deep Field, as it is the only one with a suitable quasar. Happily, results so far show that the presence of this source does not seem to be a major impediment to obtaining high dynamic range.

The significance of the radio observations is twofold.
- At levels of a few mJy the radio source population ceases to be dominated by classical radio galaxies. Instead, the radio source counts start to be dominated by weak, intermediate-redshift, young galaxies which are dominated by star formation [4]. However, fully understanding this phenomenon requires adequate sampling of this population, and this involves much more sensitive observations, extending to the limits of current instrumentation. The combination of the deep radio observations, the deep Hubble observations, and the deep observations at other wavelengths, provides a remarkable opportunity to study how these galaxies evolve to produce the relatively quiet galaxies that we see in the local Universe.
- Source count statistics suggest that amongst the objects in the HDF-S will be a handful of detectable radio sources. Indeed, our observations so far (see below) suggest that there may be rather more of these than we expect. For successful identification and classification of the objects found by the HST in the HDF-S, it is essential that we have the deepest possible radio image. In addition, HDF-S differs from the HDF-N in that there is a bright quasar nearby, and significant effort will go into identify the absorption lines and other effects from line-of-sight material. The radio studies are likely to be invaluable in identifying some of the galaxies responsible for line-of-sight absorption.

## 2  Observations

We have used the Australia Telescope Compact Array in a variety of configurations to make synthesis images of the HDF-S at four wavelengths (20, 13, 6, and 3 cm) with corresponding resolutions of about 6, 4, 2, and 1 arcsec. Our first deep radio observations started in May 1998 and observations are still continuing. So far, a total of about 400 hours have been spent observing this field, and we hope to reach a limiting rms sensitivity of about 10 $\mu$Jy with these data.

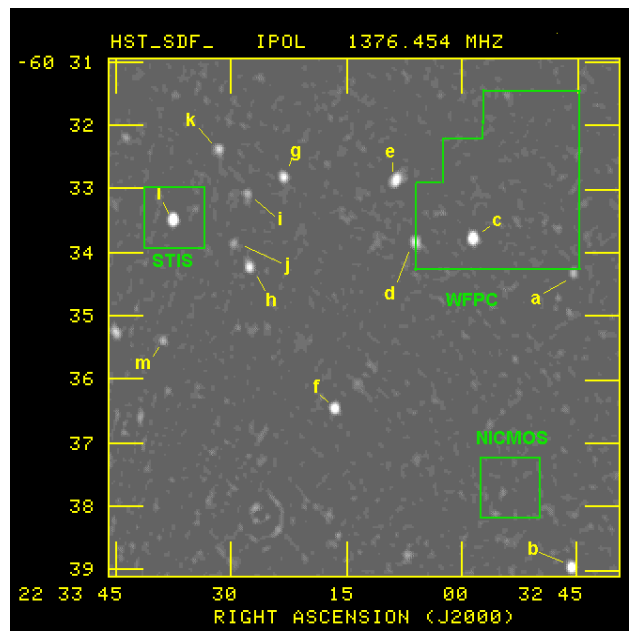

Fig 1: 20-cm ATCA image of the HDF-S. Letters indicate the 13 sources stronger than 100$\mu$Jy, and the primary WFPC, STIS, and NICMOS fields of the HST are shown.

In an analysis of about half of these data, we have imaged a region roughly 60 arcmin in extent surrounding the HDF-S at 20 cm, with correspondingly smaller regions at 13, 6, and 3 cm, and reached an rms of 18 microJy/beam. Setting a conservative detection limit of 100 microJy at 20cm, we detected about 240 sources (or 13 in the central 8 arcmin square, which contains the WFPC, STIS and NICMOS fields).

## 3  Results

Our results so far, including FITS files, "clickable" maps, and high resolution images of individual sources, are all in the public domain on http://www.atnf.csiro.au/~rnorris/hdf-s. Fig. 1 shows a 20-cm image of the 8 arcmin square containing the HDF-S and flanking fields, and identifies the 13 sources greater than 100 microJy within this field. Comparison of this image with the optical HDF-S image shows little similarity. Most strong optical sources are not strong radio sources, and vice versa. Of these 13 sources, one is the STIS quasar, eight are identified with galaxies, and four have no identified optical or infrared counterpart, indicating that R ≥ 25. The overall radio properties of this sample are statistically indistinguishable from those found in the HDF-N.

However, the most intriguing result to come from this work is the discovery of an optically faint galaxy (source 'c' in Fig. 1) with about 1000 times more radio emission than expected from conventional starburst galaxies, but much fainter optically than conventional radio galaxies.

## 4   The mysterious "source c"

Source "c" (HDFS_J223258.60-603346.6) is a 1-mJy radio source with a radio spectral index of about -0.6, and it appears to be slightly extended in the 3-cm image. Optically it appears to be a faint (R=27.16) red galaxy with B-V = 0.52 and J-K = 1.13, and appears slightly extended in the WFPC image. Fig. 2 shows this object at both radio and optical wavelengths. This galaxy is unusual in that the radio-optical ratio of this galaxy is about 1000 times higher than any object known in the local Universe.

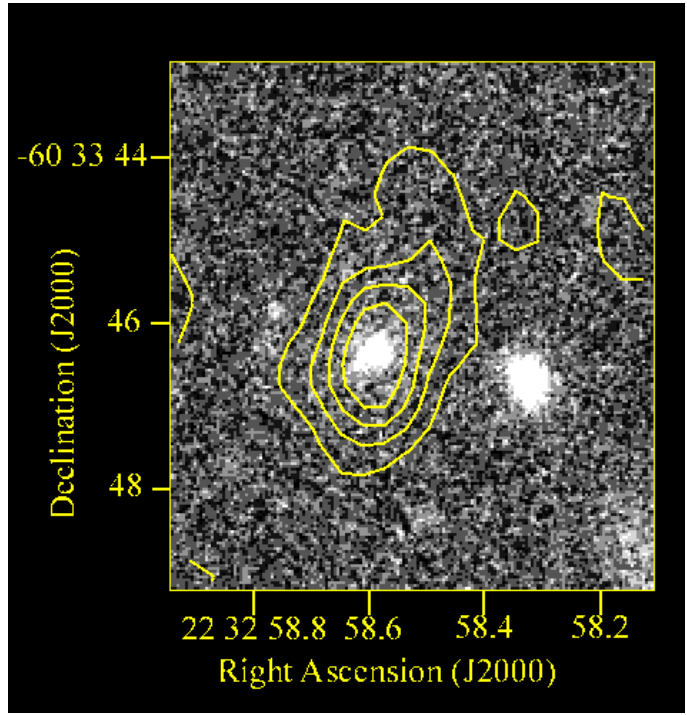

Fig 2. The 3-cm ATCA image of source c (HDFS_J223258.60-603346.6) overlaid on the 814nm WFPC image.

Because of its faintness, no spectroscopic redshift is available. Two groups have measured a photometric redshift for this galaxy. Fernandez-Soto et al. [5] have fitted an E/S0 template spectrum to the HST UBVI data and the VLT BVIJHK data (from the EIS survey) and have obtained a redshift of 1.69. However, Gwyn et al. [6] have fitted a template to the HST UBVI data alone and obtain a photometric redshift of 4.57.

Fig. 3 shows a plot of radio luminosity against optical luminosity for a number of classes of galaxy. In the lower part of the diagram are plotted the values for source c for redshifts of 1.69 and 4.71 (shown as $c_1$ and $c_2$ respectively). It should be noted that $c_1$ lies well away from the line, indicating that this is a type of galaxy not seen in the local Universe. For example, it is proportionally about 1200 times fainter at V wavelengths than Arp220 (i.e. the radio/optical ratio of this object is 1200 times that of Arp220).

Given the more extensive data used in the Fernando-Soto redshift determination, it would be natural to place more reliance on the their redshift of 1.69. However, at this redshift the radio/optical ratio is quite unlike any galaxy found in the local Universe, as shown by Fig. 3. Thus, either this object represents a previously unknown class of galaxy, or else it has a much higher redshift, which would place it closer to normal radio galaxies. At a redshift of 4.57 it would be unusual but not unique. At a redshift of 10 it would be a fairly normal radio galaxy, but would then have a Lyman break at about 1 micron, which appears to be inconsistent with the observed data.

It is now well-established that photometric redshifts tend to be reliable, and deep spectroscopic redshifts of galaxies which previously had photometric redshifts have nearly always confirmed the photometric redshifts. The discrepancy between the two photometric redshifts may indicate that this galaxy does not fit a normal template. We have not yet been able to obtain a spectroscopic redshift, which will be crucial in establishing the nature of this object.

We note that Waddington et al. [7] have discovered a faint radio galaxy (VLA_J123642+621331) in the HDF-N at a redshift of 4.424. This object is not quite so extreme as source c, since the 3 cm flux is about one third that of source c, whilst their 814nm magnitudes are similar. Nevertheless its radio/optical ratio appears to be comparable to that of source c, and it would appear just above "$c_2$" in Fig.3. Waddington et al. interpret this source as an active nucleus embedded in a starburst galaxy. Such an interpretation would also be consistent with the properties of source c.

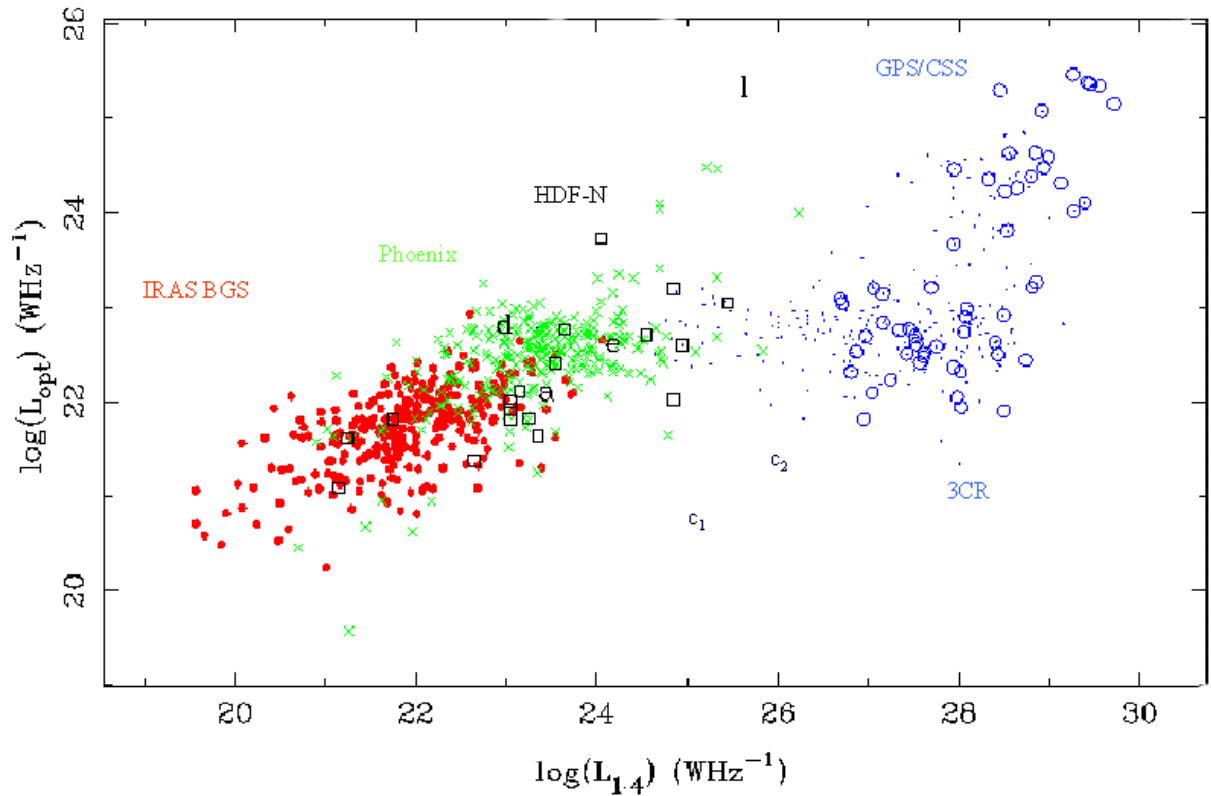

Fig 3. Plot of optical against radio luminosity for a number of samples of galaxies. Galaxies in the HDF-N with measured redshifts are shown as squares, and other galaxies from the HDF-S are shown as letters. $c_1$ and $c_2$ in the lower part of the plot show the position of source c for redshifts of 1.69 and 4.57 respectively.

## 3  Conclusion

Our preliminary radio images of the Hubble Deep Field South have uncovered a number of radio sources associated with faint galaxies. One of these, source c, remains something of a mystery. If the photometric redshift measured by Fernandez-Soto et al turns out to be correct, then this may represent a new class of galaxy, which is rare or unknown in the local Universe. We note that there are no such objects so far identified in the HDF-N, although there are a number of radio objects that correspond to blank fields at optical and infrared wavelengths, and we may speculate that some of these may be similar to source c. On the other hand, if the redshift is much higher (~5), then source c resembles VLA_J123642+621331, and is not very different from the optically faintest 3C objects. In this case it may be a radio-luminous active nucleus embedded in a very dusty starburst galaxy.

**Acknowledgements**

I thank Alberto Fernandez-Soto and Stephen Gwyn for useful comments and for making their photometric redshift data available in advance of publication.## References

[1] Fomalont, E. B.; Kellermann, K. I.; Richards, E. A.; Windhorst, R. A.; Patridge, R. B, 1997, ApJ, 475, L5.
[2] Richards, E. A.; Kellermann, K. I.; Fomalont, E. B.; Windhorst, R. A.; Partridge, R. B., 1998, AJ, 116, 1039.
[3] Williams, R.E., et al. 1998, AAS, 193, 7501.
[4] Hopkins, A., Afonso, J., Cram, L., Mobasher, B. 1999, ApJ, 519, L59
[5] Fernandez-Soto, A, private communication - see
    http://www.ess.sunysb.edu/astro/hdfs/wfpc2/html/obj0348.html
[6] Gwyn, S., private communication - see
    http://astrowww.phys.uvic.ca/grads/gwyn/pz/hdfs/pages/280.html)
[7] Waddington, I., Windhorst, R.A., Cohen, S.H., 1999, ApJL, in press.
5